\documentclass[final,5p,times,twoside,twocolumn]{elsarticle}
\pdfoutput=1
\usepackage{amssymb,amsmath}
\biboptions{sort&compress}
\usepackage{hyperref}
\usepackage{graphicx}

\newcommand{\Eq}[1]{\mbox{Eq.\,\eqref{#1}}}
\newcommand{\Fig}[1]{\mbox{Fig.\,\ref{#1}}}
\newcommand{\Sec}[1]{\mbox{Sect.\,\ref{#1}}}

\newcommand{\fc}{$f\!c$}
\newcommand{\lc}{$\ell c$}
\newcommand{\bc}{$bc$}
\newcommand{\hc}{$hc$}
\newcommand{\slc}{$s\ell c$}
\newcommand{\shc}{$shc$}

\begin{document}

\newcommand{\preprintnumbers}{\setlength{\unitlength}{1mm}
\begin{picture}(0,0)
  \put(165,20){\small HU-EP-12/43}
\end{picture}}

\begin{frontmatter}

\title{\preprintnumbers
Lattice evidence for the family of decoupling solutions of Landau gauge
Yang-Mills theory}

\date{August 2, 2013}

\author[ur]{Andr\'e Sternbeck\corref{cor}}\ead{andre.sternbeck@ur.de}
\author[hu]{Michael M\"uller-Preussker}

\cortext[cor]{Corresponding author}

\address[ur]{Institut f\"ur Theoretische Physik, Universit\"at Regensburg,  
        93040 Regensburg, Germany}
\address[hu]{Institut f\"ur Physik, Humboldt-Universit\"at zu Berlin,
  12489 Berlin, Germany}

\begin{abstract}
  We show that the low-momentum behavior of the lattice Landau-gauge gluon
  and ghost propagators is sensitive to the lowest non-trivial eigenvalue
  ($\lambda_1$) of the Faddeev-Popov operator. If the gauge fixing favors Gribov
  copies with small $\lambda_1$ the ghost dressing function rises
  more rapidly towards zero momentum than on copies with large $\lambda_1$. 
  This effect is seen for momenta below 1\,GeV, and interestingly also
  for the gluon propagator at momenta below 0.2\,GeV: For large $\lambda_1$ the
  gluon propagator levels out to a lower value at zero momentum than for small
  $\lambda_1$. For momenta above 1\,GeV no dependence on Gribov copies
  is seen. Although our data is only for a single lattice size and spacing, 
  a comparison to the corresponding (decoupling) solutions from the
  DSE/FRGE study of Fischer, Maas and Pawlowski [Annals  of Physics 324 (2009)
  2408] yields already a good qualitative agreement.
 \end{abstract}

\begin{keyword}
  Landau gauge\sep gluon and ghost propagators\sep Gribov ambiguity\sep
  Faddeev-Popov eigenvalues
\end{keyword}

\end{frontmatter}

\section{Introduction}
\label{sec:intro}

Lattice calculations of the Landau-gauge gluon, ghost and quark propagators
have attracted quite some interest during the last 15 years. Staunch
supporters of pure lattice QCD (LQCD) may wonder about the enthusiasm with
which such calculations have been performed and discussed in the past, in
particular, as LQCD comes with the distinct advantage that one does not need to
fix a gauge. This holds true, however, only as long as one is interested in
gauge-invariant quantities. But besides LQCD there are also other (sometimes
better suited) frameworks to tackle nonperturbative problems of QCD, and these
require the exact knowledge of QCD's elementary two and three-point functions in
Landau or other gauges.

Two continuum functional methods one has to mention here are the efforts to
solve the infinite tower of Dyson-Schwinger equations (DSEs) of QCD or,
likewise, the corresponding Functional Renormalization Group Equations (FRGEs)
(see, e.g., the reviews
\cite{Alkofer:2000wg,Pawlowski:2005xe,Gies:2006wv,Fischer:2006ub,Roberts:2007jh,
Maas:2011se,Boucaud:2011ug,Roberts:2012sv} and references therein). Both these
methods imply fixing a gauge
(and often the Landau gauge is chosen for simplicity), but more importantly,
these methods also require a truncation of the infinite system of equations to
enable finding a numerical solution. These truncations are a potential source of
error, which why corresponding (volume and continuum extrapolated) lattice
results are so essential to render these truncations harmless or to even
substitute parts of the DSE (or FRGE) solutions by (interpolated)
nonperturbative data. 

In what concerns the Landau-gauge gluon and ghost propagators, lattice results
have helped much to improve truncations over the years. Currently, the continuum
and lattice results overlap for a wide range of momenta, showing nice
consistency among the so different approaches to QCD. Admittedly, the currently
used truncations are still not perfect, as seen, for example, for the gluon
propagator whose DSE solutions differ from the corresponding lattice or FRGE
results in the intermediate momentum regime (i.e., for momenta $0.5-3$\,GeV),
whereas FRGE and lattice results agree much better there (see, e.g., Fig.\,2 in
\cite{Fischer:2008yv}). But this situation will certainly improve, as it did in
the past (see, e.g., \cite{Huber:2012kd} for recent progress).

Another regime that remains to be fully settled yet is the low (infrared)
momentum regime. About the infrared behavior of the gluon and ghost propagators
in Landau gauge there has been much dissent in the community and it is difficult
to assess on the lattice also. Currently, all lattice studies agree upon a gluon
propagator and ghost dressing function which are (most likely) finite in the
zero-momentum limit (see, e.g.,
\cite{Sternbeck:2005tk,Boucaud:2005ce,Cucchieri:2007md,
Sternbeck:2007ug, Bornyakov:2008yx,Bogolubsky:2009dc,
Oliveira:2008uf,Pawlowski:2009iv,Oliveira:2012eh})\footnote{Current lattice
results for this regime are for finite lattice spacings and volumes only, and
also the Gribov problem is only partially understood.}. DSE and FRGE studies
\cite{Lerche:2002ep,Zwanziger:2001kw,Fischer:2008uz,LlanesEstrada:2012my}, on
the other hand, assert that this infrared behavior is not unique, but depends
on an additional (boundary) condition on the ghost dressing function at zero
momentum, $J(0)$. Explicitly, in
Refs.~\cite{Fischer:2008uz,LlanesEstrada:2012my}, it is shown that for
$J^{-1}(0)=0$ one finds the so-called \emph{scaling} behavior for the gluon and
ghost propagators at low momentum, as it was first found in
\cite{vonSmekal:1997is}, while for finite $J(0)$, one finds a family of
\emph{decoupling} solutions for the DSEs and FRGEs, in qualitative agreement
with DSE solutions proposed in the studies of
Refs.~\cite{Aguilar:2004sw,Boucaud:2006if,Dudal:2007cw,Aguilar:2008xm,
Boucaud:2008ji,Boucaud:2008ky}, and with lattice results. For momenta above
1\,GeV both types of solutions are practically indistinguishable.
We interpret this ambiguity in the infrared as a remnant of the Gribov
ambiguity of the Landau gauge condition, which is lifted by fixing $J^{-1}(0)$
to a constant.

In this letter we will show that a part of this one-parameter family of
decoupling solutions can be seen on the lattice, at least qualitatively
and as far as it is possible on a finite and rather coarse lattice. Our
approach also allows only for mild variations of the gluon and ghost
propagators. Nonetheless, after outlining some technical details in the
next section and a discussion about the distribution of the lowest non-trivial
eigenvalue of the Faddeev-Popov (FP) operator, $\lambda_1$, on different Gribov
copies (\Sec{sec:lambda1}), we will demonstrate in \Sec{sec:results}
that the decoupling-like behavior of the lattice gluon and ghost propagators can
be changed by a (yet simple-minded implementation of a) constraint on
$\lambda_1$. Changes take place only in the low-momentum regime, but
interestingly in a similar manner as one expects from the DSE/FRGE study
\cite{Fischer:2008uz}, where a condition on $J(0)$ was used to change the
low-momentum  behavior\footnote{Note that in \cite{Fischer:2008uz} the ghost
dressing function is denoted $G$.}. Specifically, we show that on Gribov copies
with small $\lambda_1$ the ghost dressing function at low momenta rises more
rapidly towards zero momentum than on copies with large $\lambda_1$.
Interestingly, a similar (though less pronounced) Gribov-copy effect is seen for
the gluon propagator at low momentum. Qualitatively, our data thus resembles the
change of the gluon and ghost dressing functions as expected from
\cite{Fischer:2008uz} for the corresponding decoupling solutions.\footnote{We
thank C.~Fischer for providing us access to their (decoupling) solutions
including those for smaller $J(0)$ not shown in \cite{Fischer:2008uz}.}

Note that we still find Gribov copies by a maximization of the lattice
Landau-gauge functional, but we are not interested in finding Gribov copies with
large gauge-functional values, but on copies with comparably small
(or large) $\lambda_1$, irrespective of the functional value. On Gribov copies
with large gauge-functional values we see both propagators to rise less
rapidly towards zero momentum, consistent with what was found in the past
\cite{Bakeev:2003rr,Sternbeck:2005tk,Bogolubsky:2007bw,
Bornyakov:2009ug,Bogolubsky:2009qb}

We should also mention here that similar effects were seen for the
$B$-gauges by Maas \cite{Maas:2009se}. For these gauges, one selects
Gribov copies based on the ratio of the ghost dressing function at a small and a
large lattice momentum on a particular copy. By construction the ghost dressing
function in these gauges is then clearly enhanced or suppressed at low momenta.
It remains to be seen if corresponding effects become clear also for the gluon
propagator. The current data suggests, also this approach may reproduce a part
of the family of decoupling solutions on the lattice
\cite{Maas:PrivateStGoar13,Maas:2013vd}.

\section{Simulation details}
\label{sec:simulation_details}

Our study is based on 80 thermalized gauge field configurations, generated with
the usual heatbath thermalization and Wilson's plaquette action for SU(2)
lattice gauge theory. The lattice size is $56^4$ and the coupling
parameter $\beta=2.3$. To reduce autocorrelations, configurations are
separated by 2000 thermalization steps, each involving four over-relaxation and
one heatbath step. For every configuration there are at least
$N_{\mathrm{copy}}=210$ gauge-fixed (Gribov) copies, all fixed to lattice
Landau gauge using an optimally-tuned over-relaxation algorithm for the gauge
fixing that finds local maxima of the lattice Landau gauge functional
\begin{equation}
 F_U[g] = \frac{1}{4V}\sum_{x}\sum_{\mu=1}^4\,
   \operatorname{Tr}g_x U_{x\mu} g^\dagger_{x+\hat{\mu}}\,.
\label{eq:Func}
\end{equation}
Here $U\equiv\{U_{x\hat{\mu}}\}$ denotes the gauge configuration and
$g\equiv\{g_{x}\}$ one of the many gauge transformation fields fixing
$U$ to Landau gauge. To ensure these Gribov copies are all distinct, the
gauge-fixing algorithm always started from a random gauge transformation field.
Interestingly, for all these $80\times210$ gauge-fixing attempts only a few
Gribov copies were found twice.

For every single Gribov copy we determine the lowest three (non-trivial)
eigenvalues $0<\lambda_1<\lambda_2<\lambda_3$ of the Faddeev-Popov (FP) operator
using PARPACK \cite{Arpack}. In what follows, we will use $\lambda_1$ to
classify copies: The Gribov copy with lowest $\lambda_1$
(considered for each configuration separately) is labeled
\emph{lowest copy} (\lc), while the copy with the highest $\lambda_1$ we call
\emph{highest copy} (\hc). The first generated copy, irrespective of
$\lambda_1$, gets the label \emph{first copy} (\fc). It represents an
arbitrary (random) Gribov copy of a configuration. To compare with
former lattice studies on the problem of Gribov copies we also reintroduce the
label \emph{best copy} (\bc). It refers to that copy with the best (largest)
gauge functional value $F_U[g]$ for a particular gauge configuration.

On those four sets of Gribov copies we calculate the SU(2) gluon and ghost
propagators following standard recipes. That is, the gluon propagator is
calculated for every lattice momentum using a fast Fourier transformation and
the ghost propagator by using the plane-wave method for selected momenta. To
accelerate the latter we use the preconditioned conjugate
gradient algorithm of \cite{Sternbeck:2005tk}. As a by-product of
this calculation we also obtain the renormalization constant, $\widetilde{Z}_1$,
of the ghost-ghost-gluon (gh-gl) vertex in Landau gauge for zero incoming gluon
momentum. For more details on lattice Landau gauge and the
calculation of the propagators and $\widetilde{Z}_1$ the reader may refer
to Refs.~\cite{Cucchieri:2004sq,Sternbeck:2005tk,Sternbeck:2006rd,Sternbeck:2008mv} 
and references therein.

When quoting momenta in physical units we adopt the usual definition
$ap_{\mu}(k_{\mu}) = 2 \sin\left(\pi k_{\mu}/L_{\mu}\right)$ with
$k_{\mu}\in(-L_{\mu}/2, L_{\mu}/2]$ and $L_{\mu}\equiv56$, assume
for the string tension
$\sqrt{\sigma}=440\,\textrm{MeV}$ and use $\sigma a^2 = 0.145$ for $\beta=2.3$
from Ref.~\cite{Langfeld:2007zw}, where $a$ denotes the lattice spacing.

\section{Distribution of $\lambda_1$}
\label{sec:lambda1}

\begin{figure}[t]
  \centering
  \includegraphics[width=0.95\linewidth]{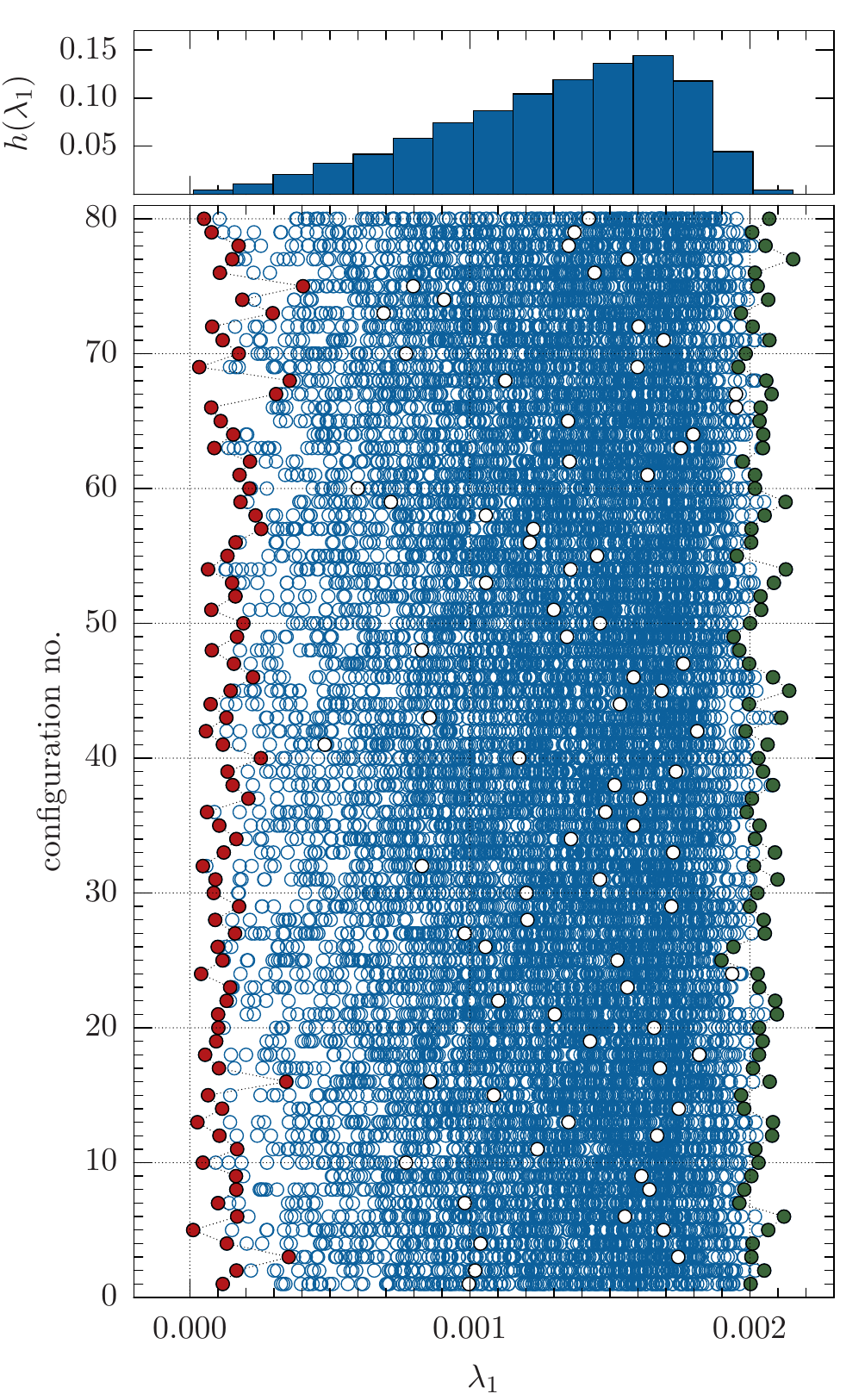}
  \caption{Distribution of $\lambda_1$ (in lattice units) for 210 Gribov
    copies, shown separately for 80 thermalized gauge configurations (ordinate).
    Full red (white, green) circles mark the eigenvalue on \lc\ (\fc, \hc)
    copies. The small top panel shows as a histogram for $\lambda_1$ including
    all values found.} 
\label{fig:eigenvalues_vs_conf}
\end{figure}

Before comparing the propagator data for the different types of Gribov copies,
it is instructive to look at the distribution of $\lambda_1$ on all copies
first. In \Fig{fig:eigenvalues_vs_conf} we show this eigenvalue distribution
(in lattice units) for $N_{\mathrm{cp}}=210$ Gribov copies. There, the big
panel shows it separately for each of the 80 gauge configurations and the small
panel (on top) for all configurations together as a histogram. One sees that for
most of the copies $\lambda_1$ takes values between $0.5\times 10^{-3}$ and
$1.9\times 10^{-3}$, mostly between $1.5\times 10^{-3}$ and $1.7\times
10^{-3}$, but for some configurations there are also copies with an
exceptionally small value for $\lambda_1$, a value ($\lambda_1<10^{-4}$) far
below the values found for the other copies. With our simple (brute-force)
approach we are rather limited in finding more of these exceptional copies. The
gauge-fixing and calculation of eigenvalues on a $56^4$ lattice is computational
quite demanding, and a more sophisticated gauge-fixing algorithm---one which
would automatically select that Gribov copy with the smallest (or at least with
small) $\lambda_1$---does not exist. But it would be interesting to know if for
each configuration a Gribov copy with such an exceptionally small $\lambda_1$
exists. 

For a few configurations we generated more than 210 Gribov copies. These allow
us now to have a closer look at the distribution of~$\lambda_1$ and to
demonstrate that this number of copies is sufficient to resemble the
distribution's shape for our lattice parameters ($\beta=2.3$,
$56^4$). 

\begin{figure*}
 \mbox{\includegraphics[height=7.5cm]{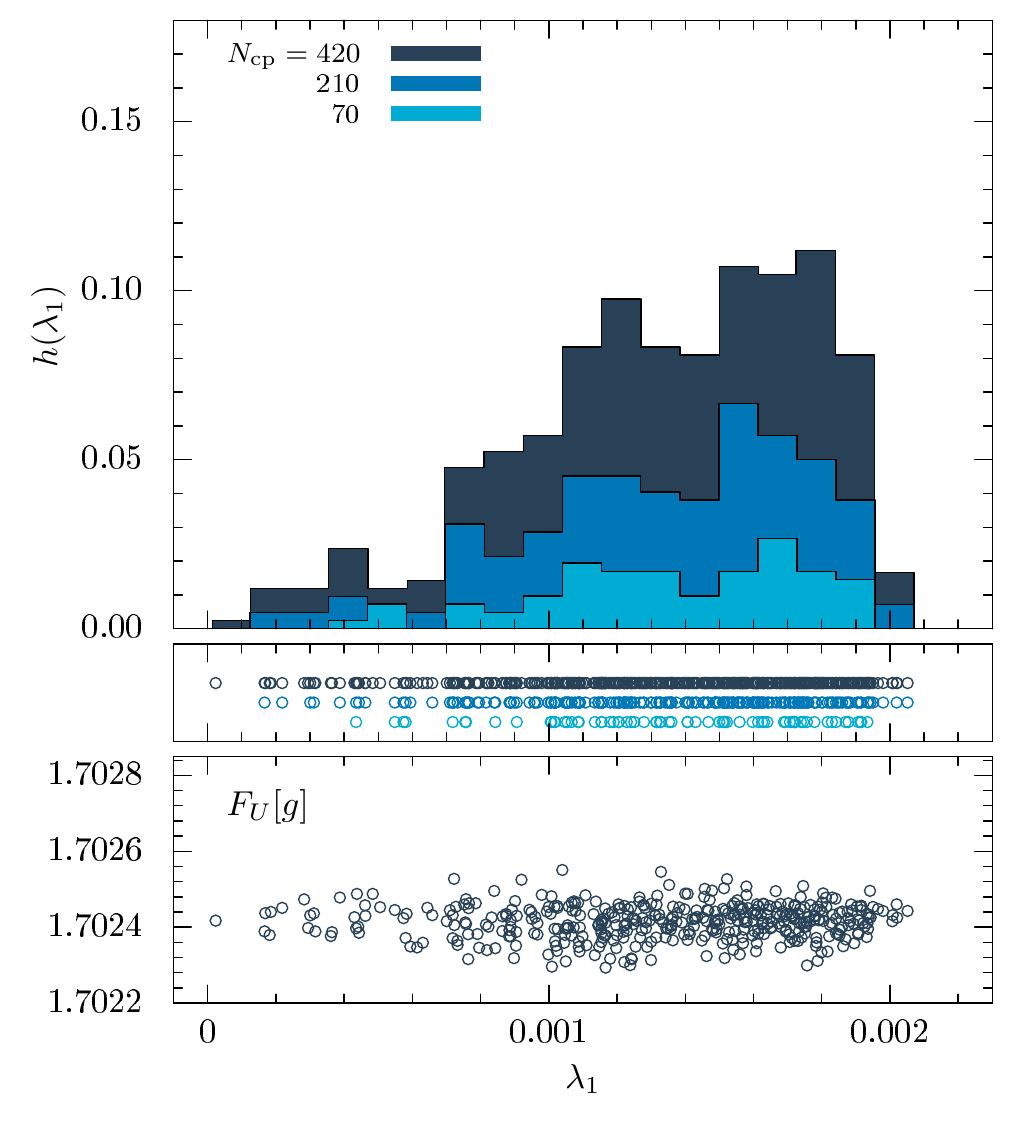}\hspace{-2ex}
 \includegraphics[height=7.5cm]{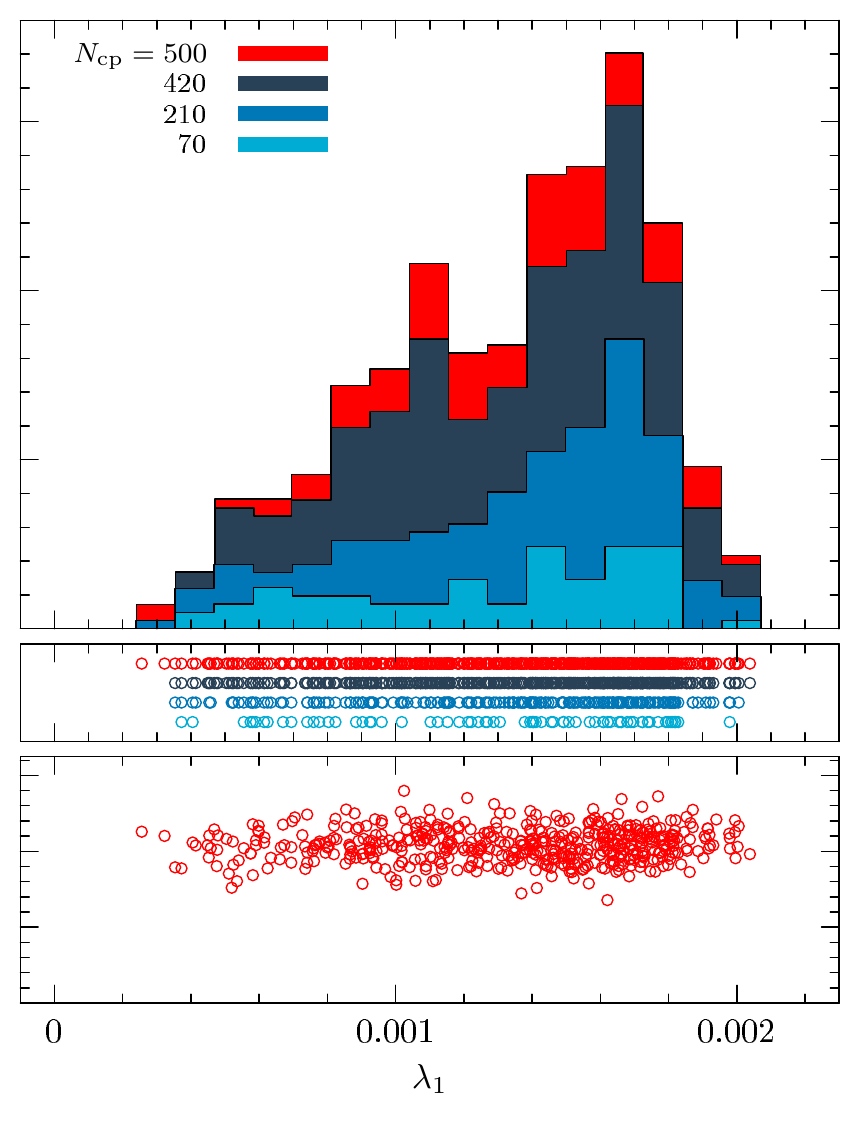}\hspace{-2ex}
 \includegraphics[height=7.5cm]{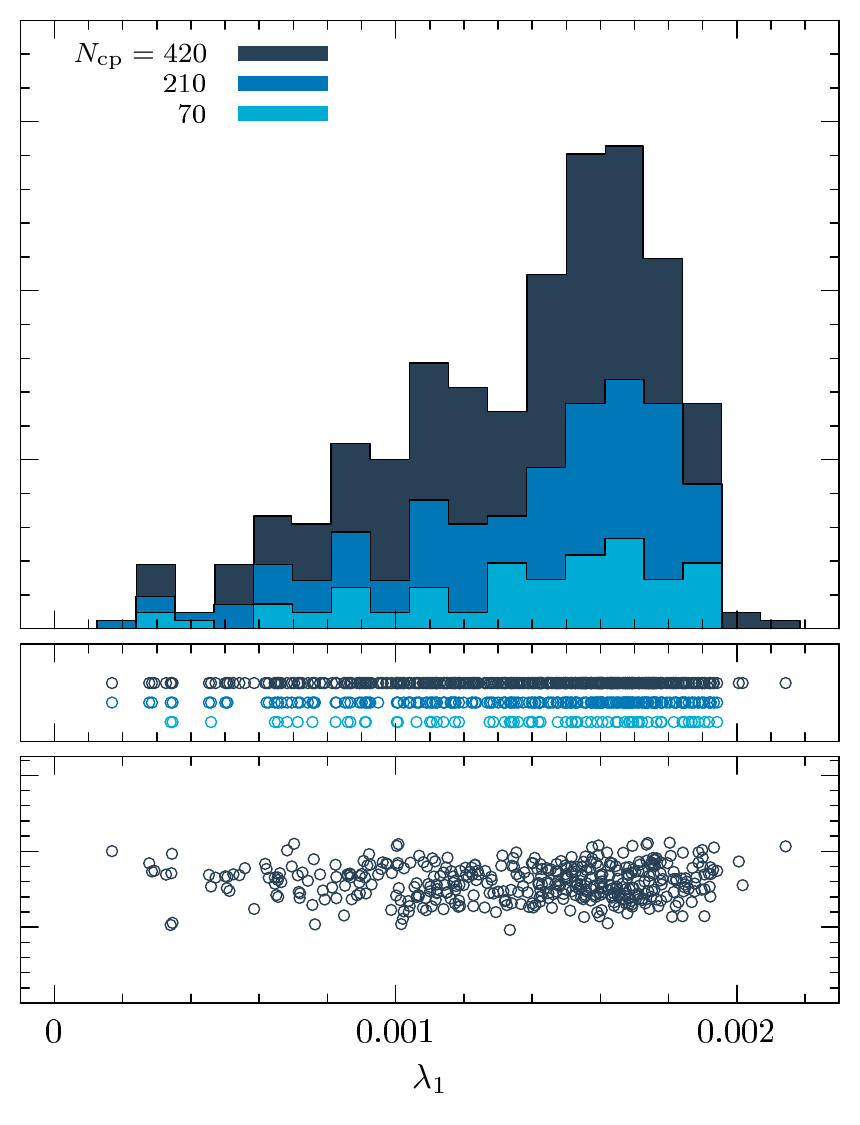}}
 \caption{Top panels: histogram of the eigenvalue distribution of
     $\lambda_1$ for different numbers $N_{\mathrm{cp}}$ of Gribov
     copies. From left to right, each panel is for one gauge configuration.
      All histograms are normalized such that the sum
      of bars is one for $N_{\mathrm{cp}}=420$ copies. 
      The middle panels show the corresponding scattering of $\lambda_1$ for
the different $N_{\mathrm{cp}}$, and the lower panels show scatter plots of
$\lambda_1$ versus the gauge functional values $F_U[g]$ for the
respective largest $N_{\mathrm{cp}}$.} 
\label{fig:evdist_lambda1}
\end{figure*}

Typical snapshots of this distribution for different $N_{\mathrm{cp}}$
are shown in \Fig{fig:evdist_lambda1}, from left to right for three arbitrarily
selected configurations. There, the middle panels illustrate how the range of
$\lambda_1$-values gets populated when increasing $N_{\mathrm{cp}}$; shown are
distributions for $N_{\mathrm{cp}}=70$, 210, 420 and 500 Gribov copies. The top
panels show the corresponding histograms, filled with the respective symbol
color used in the middle panels. To ease the comparison, these histograms are
all normalized with respect to $N_{\mathrm{cp}}=420$.

From these histograms we see that the individual (configuration-wise)
distributions of $\lambda_1$ are asymmetric and negatively skewed, similar to
what we have just seen for the overall distribution in
\Fig{fig:eigenvalues_vs_conf}. Moreover, we see that at least 200 Gribov
copies are needed to reach at an approximate shape for the distribution. This
number of copies seems to be also sufficient (for the given lattice parameters)
to find copies with either very small or very large $\lambda_1$, even
though it is unlikely that these copies are those with the respective smallest
and largest $\lambda_1$ overall. 

We are thus in a good position to analyze the correlation between $\lambda_1$
and the low-momentum behavior of the gluon and ghost propagator.

The attentive reader may have noticed the bottom panels of
\Fig{fig:evdist_lambda1}. These show correlation plots of $\lambda_1$ versus the
gauge functional $F_U[g]$, always for the largest available number of Gribov
copies ($N_{\mathrm{cp}}=420$ or 500) for the respective configuration. In
former lattice studies on the infrared behavior of the gluon and ghost
propagators (e.g.,
\cite{Bogolubsky:2005wf,Bornyakov:2008yx,Bogolubsky:2009dc,Bornyakov:2009ug}),
gauge-fixing algorithms were often designed to find copies with comparably large
$F_U[g]$, in the hope these copies are closer to the fundamental modular region
than a set of random copies can be. Gribov copies which globally maximize the
gauge functional are elements of the fundamental modular region, and it was
argued \cite{Zwanziger:1993dh,Zwanziger:1998ez,Zwanziger:2003cf}
that in the continuum the common boundary of this region and the Gribov
horizon (the set of Gribov copies with $\lambda_1=0$) is expected to dominate
the path integral in the thermodynamic limit. This means that expectation
values of correlation functions integrated over the fundamental
modular region equal those integrated over the Gribov region
\cite{Zwanziger:2003cf}. It is however still open how these findings fit to the
current lattice data, because in the Gribov-Zwanziger (GZ) approach the
horizon condition also entails a gluon propagator (ghost dressing
function) which vanishes (diverges) in the zero-momentum limit, which
is not what is commonly seen on the lattice (see, e.g.,
\cite{Zwanziger:2002ia,Zwanziger:2003cf,Fischer:2008uz,Maas:2013vd}
for a more detailed discussion). These lattice results much better fit
to the findings for the Refined-GZ approach \cite{Dudal:2008sp} that
also implements the horizon condition.

It is therefore interesting to check if small values for $\lambda_1$ are
correlated with large values for $F_U[g]$. Looking at the scatter plots 
for $F_U[g]$ versus $\lambda_1$ (lower panels of
\Fig{fig:evdist_lambda1}), we find, however, there is no obvious correlation
between them. There are copies with small $\lambda_1$ and small $F_U[g]$, but
at the same time there are also copies with small $\lambda_1$ and large
$F_U[g]$, and vice versa. Though, we see that on average highest copies tend to
yield somewhat larger $F_U[g]$ values, in particular compared to random copies.
This is consistent with our findings in \cite{Sternbeck:2005vs}.

\section{$\lambda_1$ and the gluon and ghost propagators}
\label{sec:results}

Next we look at the correlation of $\lambda_1$ and the gluon and ghost
propagator at low momenta. For the gluon propagator $D(k)$ [or its
dressing function $Z(k) = p^2 D(k)$] there is actually no direct
relationship between its momentum dependence and the spectrum of the FP
operator, besides that it is gauge-dependent and so may change if the lattice
Landau gauge condition is supplemented by a condition on $\lambda_1$.

The ghost propagator $G(k)$ [or its dressing function $J(k)=p^2 G(k)$], 
on the other hand, should be affected stronger.
Given its spectral decomposition in SU(2),
\begin{equation}
  G(k) = 
\frac{1}{3}\sum_{i=1}^{3V-3}\frac{\vec\Phi_i(k)\cdot\vec\Phi_i(-k)}{\lambda_i}\,,
 \label{eq:decomposition_of_(ghost}
\end{equation} 
it is plausible that Gribov copies with very small $\lambda_1$ yield
larger values for $G$ than copies where $\lambda_1$ is comparably large, 
possibly even configuration-wise. Though this is not as
simple as it seems at first sight, because $\lambda_1$ and the corresponding
eigenfunction $\Phi_1(k)$ contribute only a minor fraction to $G(k)$ and this
fraction even shrinks the larger the lattice momentum $a^2p^2(k)$ (see
Ref.~\cite{Sternbeck:2005vs}, in particular Fig.\,6 therein). However, from
\cite{Cucchieri:2008fc} (Fig.\,1) we learn this fraction seems to increase
with volume. That is, an anti-correlation of $\lambda_1 $ and $G$ is
suggested.

For a few gauge configurations we have data for the gluon and ghost propagator
for all Gribov copies. It allows us to check if there is a direct correlation
between $\lambda_1$ and the gluon and ghost propagator at low momenta, comparing
different Gribov copies of the same configuration. Looking at this data reveals,
however, no immediate correlation: A Gribov copy with a smaller $\lambda_1$ not
necessarily yields a larger ghost propagator than a copy with a somewhat larger
$\lambda_1$. Similar we find for the gluon propagator. 

Nonetheless, when looking at these correlations more broadly, that is
on average for the whole gauge ensemble, we see a clear trend: Gribov copies
with very small $\lambda_1$ tend to yield a larger ghost propagator at low
momentum, than copies with large $\lambda_1$ (see, e.g., Fig.~2 in
\cite{Sternbeck:2013zja}). This anti-correlation is smaller the
larger the momentum, and is also considerably smaller for the gluon propagator
at same momentum. 

\begin{figure*}[!t]
 \centering
 \mbox{\includegraphics[height=15cm]{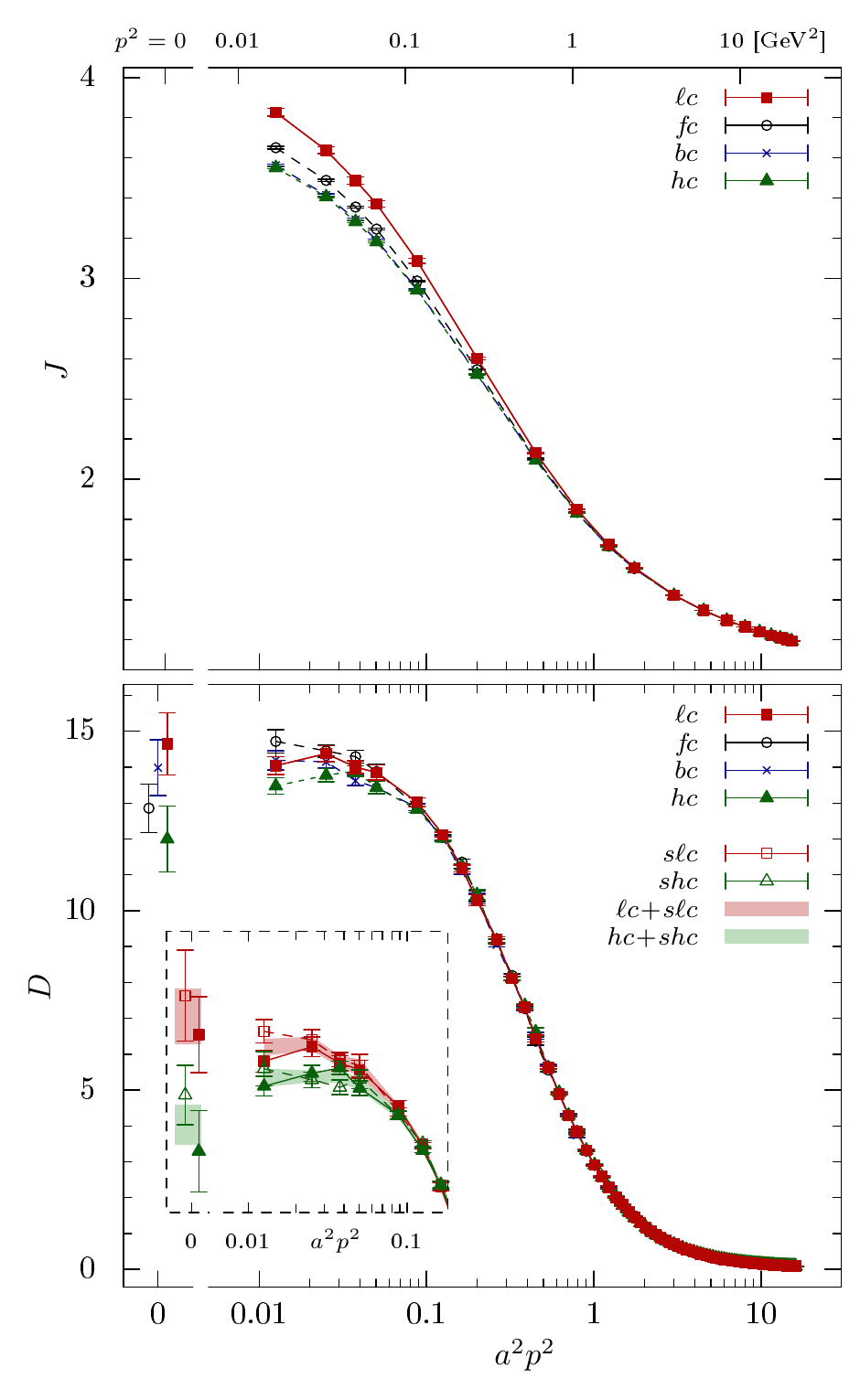}\hspace{
3ex}
 \includegraphics[height=15cm]{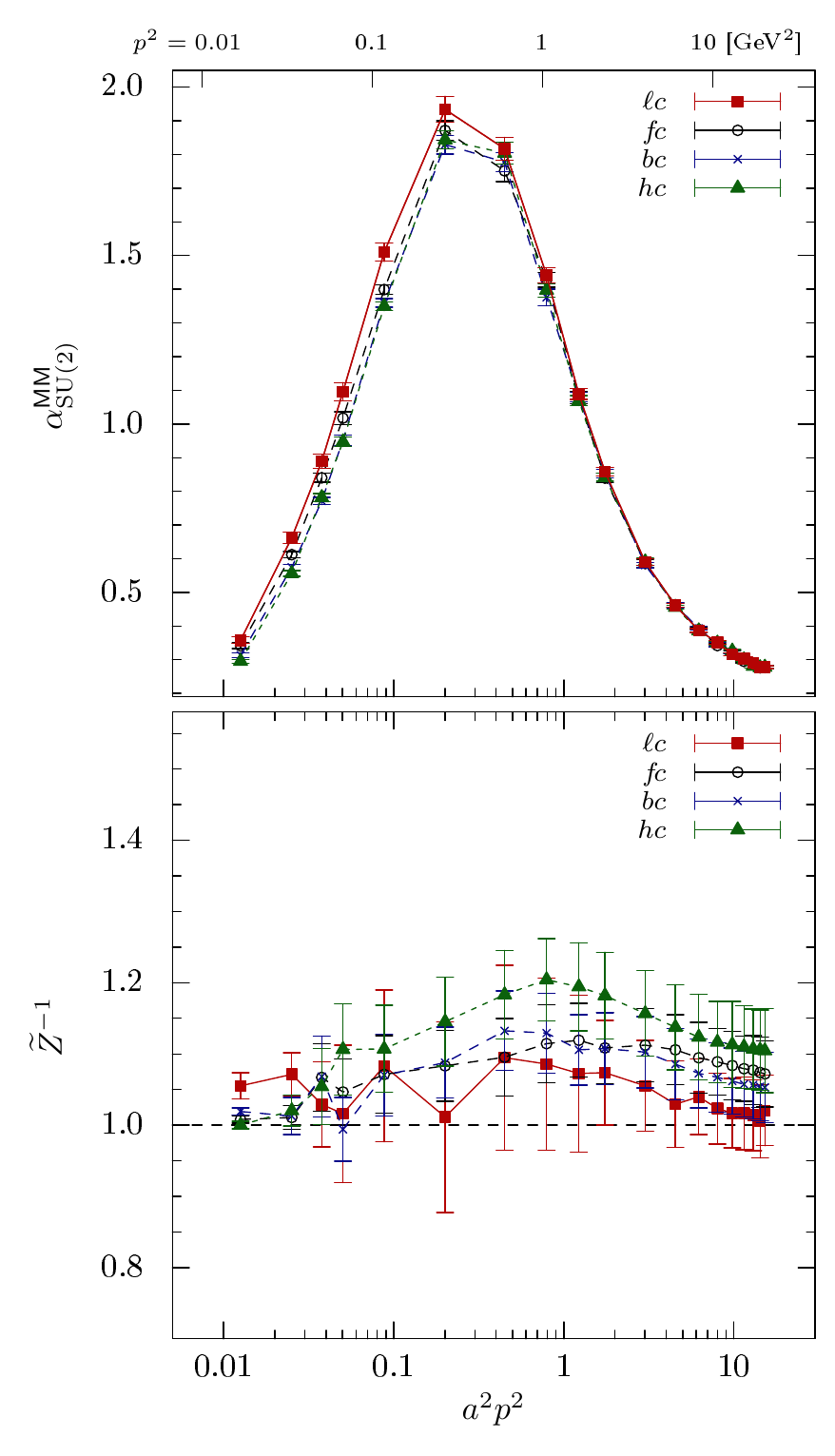}}
 \caption{Ghost dressing function (top left), gluon propagator (bottom
  left), the coupling $\alpha^{\mathsf{MM}}_{\mathrm{SU(2)}}$ (top right)
  and the inverse renormalization constant of the gh-gl-vertex (bottom right);
  all versus lattice momentum squared, in lattice units and for SU(2). No
  renormalization has been applied. Open circles (filled squares, crosses, full
  triangles) refer to \fc\ (\lc, \bc, \hc) data. The lower left panel also shows
  a zoomed-in plot to improve visibility of the low-momentum region. This
  panel also shows \slc\ and \shc\ data. Note
  the different scales for the x-axis ($\log$ and  linear) in the left panels
  and the corresponding physical momenta above the top panels. The $D(0)$
  points are slightly shifted to improve visibility.}
\label{fig:ghdress_gl_alpha_qq} 
\end{figure*}

These correlations are also seen when comparing averaged lattice data as
shown in \Fig{fig:ghdress_gl_alpha_qq}. There the left panels compare
the momentum dependence of the SU(2) ghost dressing function (top) and the gluon
propagator (bottom) as obtained on first, lowest, highest and best copies
of our gauge ensemble (refer to \Sec{sec:simulation_details} for this
classification). The right panels confront the corresponding data for the strong
coupling constant, 
\begin{equation}
 \alpha^{\mathsf{MM}}_{\mathrm{SU(2)}}(p) = \frac{g_0^2(a)}{4\pi}\, Z(a,p)
\cdot J^2(a,p)\,\,,
 \label{eq:alpha}
\end{equation} 
here in the Minimal MOM scheme for SU(2) Landau gauge \cite{vonSmekal:2009ae},
and of the inverse renormalization constant $\widetilde{Z}_1^{-1}$ of the
\mbox{gh-gl} vertex with zero incoming gluon momentum. 

Looking first at the first-copy (\fc) data points (black open circles
in \Fig{fig:ghdress_gl_alpha_qq}), we see these behave as expected
\cite{Cucchieri:2007md,Sternbeck:2007ug,Bogolubsky:2009dc,Bogolubsky:2009qb,
Bornyakov:2009ug}: The ghost dressing function and the gluon propagator
increase with decreasing momentum and tend to reach a plateau at very
low momentum, while $\alpha^{\mathsf{MM}}_{\mathrm{SU(2)}}$ grows with
decreasing momentum down to about $a^2p^2=0.2$ ($p\approx0.5$\,GeV) below which
it falls off again with momentum. Also $\widetilde{Z}_1^{-1}$ is found as
expected: $\widetilde{Z}^{-1}_1\to 1$ for large momenta, as it should,
and for intermediate momenta we see its characteristic hump
\cite{Cucchieri:2004sq,Sternbeck:2006rd,Cucchieri:2008qm}.

But the most interesting data in \Fig{fig:ghdress_gl_alpha_qq} is
the lowest (\lc) and highest-copy (\hc) data (squares and triangles). The
\lc~data points, for example, for the ghost dressing function clearly deviate
upwards from the respective \hc~data below $a^2p^2=0.5$. Similar is
seen for the gluon propagator below $a^2p^2=0.03$. These deviations then
yield a coupling, $\alpha^{\mathsf{MM}}_{\mathrm{SU(2)}}$, whose running is
slightly upwards/downwards shifted below $a^2p^2=0.5$. 

Admittedly, compared to the ghost dressing function, the effect for the gluon
propagator is small, but we find that it becomes more and more pronounced when
increasing statistics. To cross-check that the $\lambda_1$-dependence we see
for the gluon propagator is not just a statistical artifact, we performed
additional
calculations of the gluon propagator on all Gribov copies with
the second lowest and second highest $\lambda_1$. Those additional sets of
($2\times 80$) Gribov copies (labeled \slc\ and \shc\ in what follows) are
distinct from the sets of lowest and highest copies (\lc\ and \hc) analyzed
above, and if there is a dependence on $\lambda_1$, one should also see it when
comparing \slc\ and \shc\ data. And in fact, also this data clearly
exhibits a $\lambda_1$-dependence at low momenta (see
zoomed-in plot in \Fig{fig:ghdress_gl_alpha_qq}). The
combined \lc\ and \slc\ data and the combined \hc\ and \shc\ data (colored error
bands, same panel) currently gives the
best impression of this dependence. Note that such a combination of data is
justified, as there are no correlations visible between data from different
copies of the same configuration, and by construction these sets of Gribov
copies come also with similar small or large values for $\lambda_1$: The
averaged $\lambda_1$ values on these four sets of Gribov copies are: 
$\langle\lambda_1\rangle_{\ell c}=1.43(9)\times10^{-4}$,
$\langle\lambda_1\rangle_{s\ell c}=2.14(9)\times10^{-4}$,
$\langle\lambda_1\rangle_{hc}=20.33(6)\times10^{-4}$ and
$\langle\lambda_1\rangle_{shc}=19.89(4)\times10^{-4}$, all in lattice units.

So we think the effects we find for both propagators and
$\alpha^{\mathsf{MM}}_{\mathrm{SU(2)}}$ are not statistical
artifacts, but result from our selection of Gribov copies with respect to 
$\lambda_1$. A selection based on values for the ghost dressing function, as
proposed in \cite{Maas:2009se}, would probably result in something similar.

For completeness, we also show the corresponding best-copy (\bc) data
in \Fig{fig:ghdress_gl_alpha_qq} (blue crosses, partly hidden by
the \hc~data). Comparing this with
the \fc~data, we find the expected suppression of the \bc~data for the ghost and
gluon propagator at low momentum which is then also seen for
$\alpha^{\mathsf{MM}}_{\mathrm{SU(2)}}$. Similar was seen for the \bc~data in
\cite{Bornyakov:2008yx,Bornyakov:2009ug}, although there the effect was even
bigger when using the FSA gauge-fixing, a special combination of the
simulated annealing and over-relaxation algorithm for gauge-fixing and $Z(2)$
flips, which is most suitable for finding Gribov copies with large $F_U[g]$
values.

Note that the coincidence of the \hc\ and \bc\ data for the ghost dressing
functions is likely to be accidentally. For these two sets the dependence on the
gauge-functional value (decrease for increasing $F_U[g]$) adds to the
dependence on $\lambda_1$ (decrease for increasing $\lambda_1$). See, once
again, Fig.~2 in \cite{Sternbeck:2013zja}.

For all four types of Gribov copies, we find $\widetilde{Z}^{-1}_1$
remains almost unaffected (see lower right panel of
\Fig{fig:ghdress_gl_alpha_qq}). There are small upward
shifts for the \hc\ data compared to the \lc\ data, but these shifts are all
within (statistical) errors. This certainly deserves further study, because at
large momenta $\widetilde{Z}_1$ should approach $1$ for all types of copies
\cite{Taylor:1971ff,vonSmekal:2009ae}, while at small momenta
$\widetilde{Z}_1$ is expected to approach $1$, or at least a value close to
$1$ \cite{Lerche:2002ep}.

\section{Discussion and Summary}
\label{sec:conclusion}

\begin{figure}[!t]
 \centering
 \includegraphics[width=\linewidth]{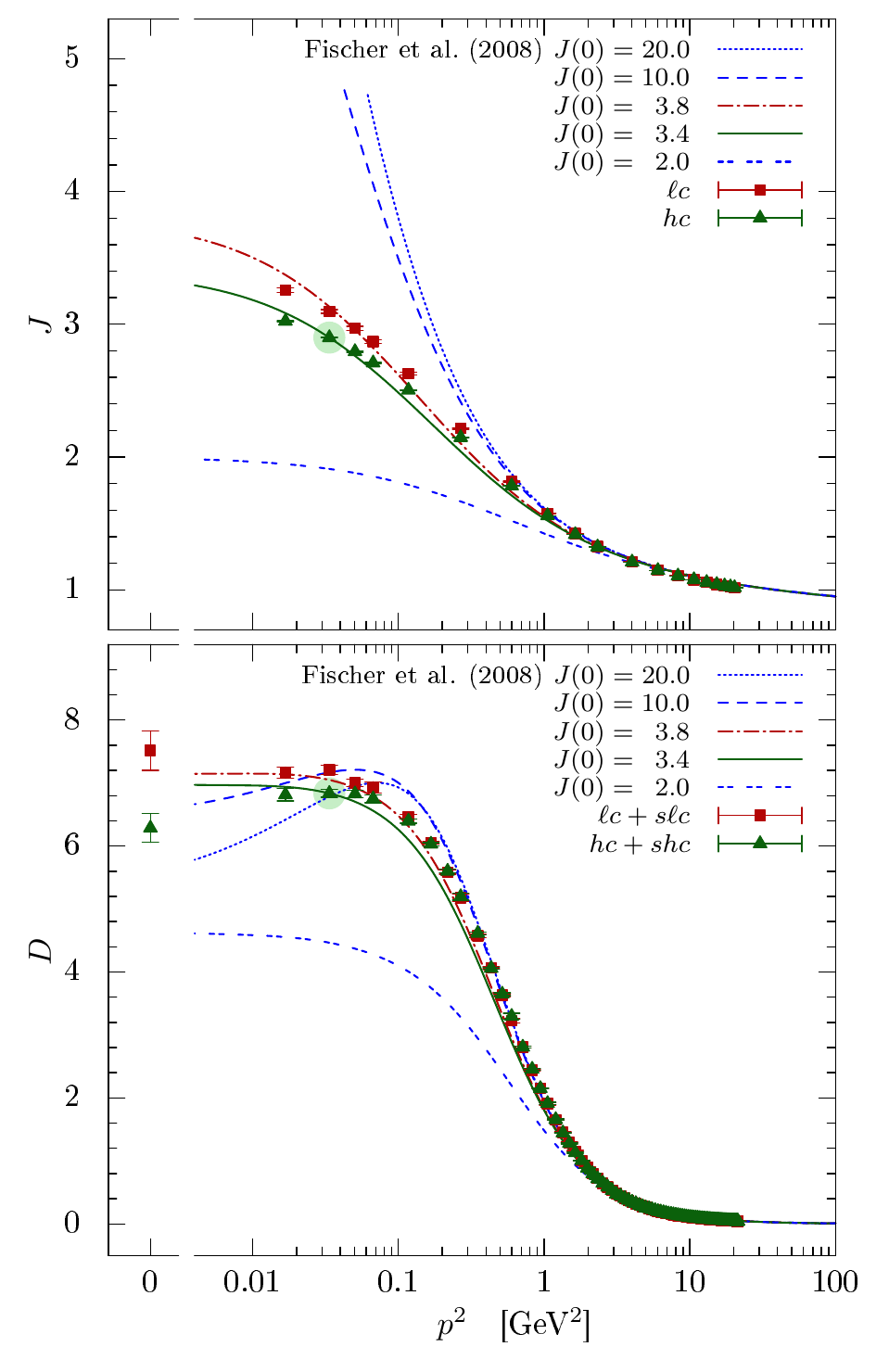}
 \caption{Ghost dressing function (top) and gluon propagator (bottom)
    versus $p^2$. Full symbols refer to our lattice data and lines to
    five selected decoupling (DSE) solutions from \cite{Fischer:2008uz}. Note,
    the order of the gluon propagator lines at low momenta changes
    somewhere between $J(0)=3.8$ and 10. The two green-highlighted triangles
    have been used to fix the relative renormalization factors $Z_J$ and $Z_D$
    between the data and the decoupling solutions (see text).}
\label{fig:gl_ghdress_qq_lc2lc_hc2hc_DSE}
\end{figure}

We have demonstrated that the low-momentum behavior of the Landau-gauge gluon
and ghost propagators can be changed on the lattice by an additional
constraint on the lowest-lying (non-trivial) eigenvalue, $\lambda_1$,
of the FP operator. If the lattice Landau gauge fixing is tuned to find
Gribov copies with a very small $\lambda_1$, the ghost dressing function
gets more enhanced towards the infrared momentum limit, while it rises less
steep for large $\lambda_1$. A similar but less pronounced effect is seen
for the gluon propagator. Currently the effect for the gluon propagator is best
seen if one combines the data from Gribov copies with lowest and second lowest
$\lambda_1$, and from copies with highest and second highest $\lambda_1$ as
shown, e.g., in Figs.~\ref{fig:ghdress_gl_alpha_qq} or
\ref{fig:gl_ghdress_qq_lc2lc_hc2hc_DSE}. The combined effect of gluon and ghost
propagator then yields a coupling constant
$\alpha^{\mathsf{MM}}_{\mathrm{SU(2)}}$ whose running changes slightly
for momenta below 0.7\,GeV.

These Gribov-copy effects are not noticeable at momenta above 1\,GeV.
Also, these effects cause only a quantitative but no qualitative
change of both these propagators at low momentum.
Their modified momentum dependences are still of decoupling type. But
interestingly, the change we find with $\lambda_1$ looks very much alike the
change of the gluon and ghost dressing functions, $Z$ and $J$, with the
(boundary) condition on $J(0)$ as found in~\cite{Fischer:2008uz}.

To demonstrate this we confront our data for the ghost dressing function and the
gluon propagator in \Fig{fig:gl_ghdress_qq_lc2lc_hc2hc_DSE} with a
corresponding pair of decoupling solutions from
\cite{Fischer:2008uz}\footnote{The DSE solutions we us here are renormalized
such that $\alpha^{\mathsf{MM}}_s(\mu)=0.3$.}. These
DSE solutions are approximately those where the boundary condition on the ghost
dressing function was set to $J(0)=3.4$ and $J(0)=3.8$, respectively. For the
comparison our data has been renormalized relatively to the given decoupling
solution by applying a common renormalization factor ($Z_J$ and $Z_D$) to the
respective data. Since DSE truncation effects on the solutions are expected to
become important for intermediate (around 1 GeV) and higher momenta, $Z_J$ and
$Z_D$ were chosen such that the \hc-data points (green triangles) agree with the
$J(0)=3.4$ curves (green) at the second lowest finite momenta (this point is
highlighted by green circle in the figure). Of course, one could choose any
other point at low $p^2$, which would result in a similar figure. But at the
moment our comparison is only qualitative anyway.  

\begin{figure}[t]
 \centering
 \includegraphics[width=\linewidth]{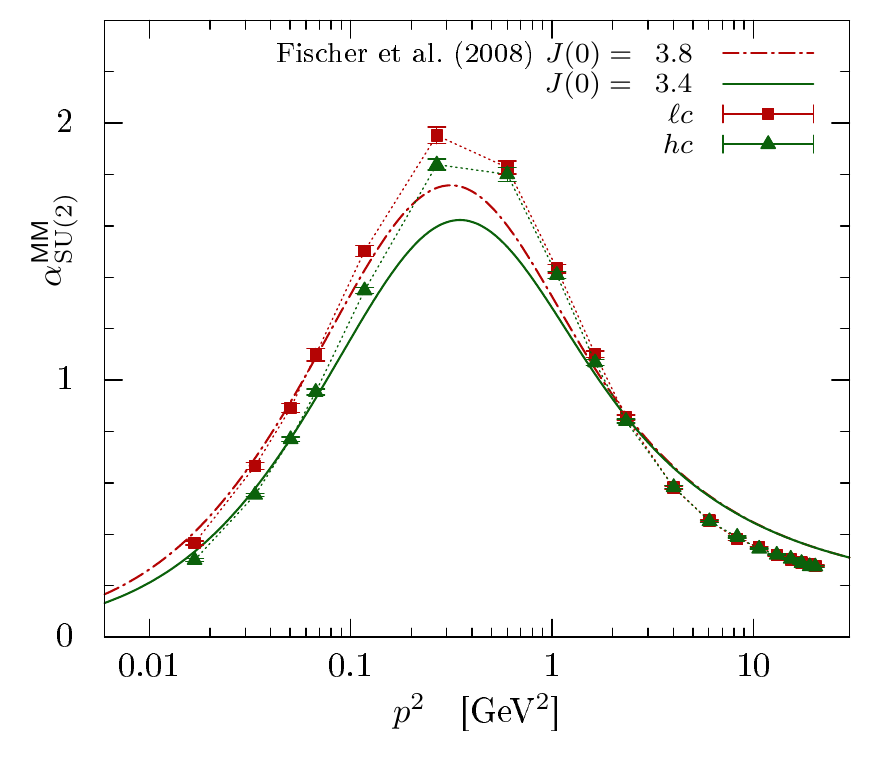}
 \caption{The strong coupling constant [\Eq{eq:alpha}] versus $p^2$. Symbols
refer to our lattice data (same as in \Fig{fig:ghdress_gl_alpha_qq})
    and (full/dash-dotted) lines to two selected decoupling solutions from
    \cite{Fischer:2008uz}.}
\label{fig:alpha_qq_DSE}
\end{figure}

A more quantitative comparison is possible though, if one looks directly at the
strong coupling constant [\Eq{eq:alpha}]. For this \emph{no} relative
renormalization factors are needed and the respective DSE and lattice data  
should agree, apart from discretization and finite-size effects (for the lattice
data) and truncation effects (for the DSE solutions) at intermediate and larger 
momenta. In \Fig{fig:alpha_qq_DSE} we show such a comparison for the decoupling
(DSE) solutions with $J(0)=3.4$ and 3.8, respectively. The surprisingly good
agreement between the propagators and the coupling constant at small momenta
for the so different approaches to Landau-gauge Yang-Mills theory is
encouraging. 

Note again that only for small momenta a quantitative agreement can
be expected at present. This is primarily due to the used DSE solutions
themselves. These were obtained for a truncated system of gluon and ghost DSEs
(see \cite{Fischer:2008uz}) and it is known that these truncations affect the
solutions, in particular at intermediate momenta. Also the running of
$\alpha^{\mathsf{MM}}_s(p)$ at very large $p$ is only exact up to 1-loop order
for these solutions \cite{Maas:PrivateStGoar13}. Improved truncation schemes
will however help to reduce the gap between lattice data and DSE solutions in
the future (see, e.g., \cite{Huber:2012kd} where an improvement has been
achieved recently). But to be fair, also our lattice data is not yet perfect:
it is for a large ($aL_\mu\approx 9.6\text{fm}$) but still rather coarse
lattice and neither infinite-volume nor continuum-extrapolated.
That is, the lattice points at larger momenta will slightly deviate from the
momentum dependence in the continuum limit.\footnote{An example of how these
deviations look like is given Fig.\,2 of \cite{Sternbeck:2012qs} for 1-loop
lattice perturbation theory.} 

It is because of the fact that all these effects are less severe for small
momenta, why a quantitative agreement should be expected only there.

Nonetheless, with our somewhat exploratory study we primarily focused on the
correlation between the FP eigenvalue spectrum and the ghost and gluon
propagators, in particular if a change in a constraint on $\lambda_1$ is
reflected in a \emph{simultaneous} change of the ghost and gluon propagators at
low momentum, and if this change qualitatively agrees with
\cite{Fischer:2008uz}. 
We therefore chose SU(2) and a fixed but rather coarse $56^4$ lattice
($\beta=2.3$) on purpose. It has allowed us to get (with reasonable computer
time) data for the gluon propagator and the ghost dressing function where their
momentum dependence starts to become flat, and this for a sufficient number of
Gribov copies (at least 200 per configuration) such that a correlation between
$\lambda_1$ and both propagators could be seen. Also, the use of a
single lattice size and spacing has the advantage that one can reveal the effect
without having to correct for other effects at the same time (finite volume,
lattice spacing, renormalization). It brings, however, the disadvantage
that we have no control yet over finite-volume and discretization effects.

For a future study, we therefore suggest to explore how the correlations
between the ghost dressing function, the gluon propagator and $\lambda_1$
change if the lattice volume is increased and the lattice spacing decreased.
This is important, in particular for the gluon propagator whose low-momentum
behavior changes only little, if one constrains $\lambda_1$, or $J(0)$ in the
continuum.
Also 3-point functions of gluon and ghost fields should be checked for this
effect, and what changes when adding fermions. Such calculations will however
become expensive very quickly---especially, if Gribov copies with exceptionally
small or large $\lambda_1$ still have to be found by chance. Algorithmic
improvements would therefore be quite helpful. For example, a gauge-fixing
algorithm that automatically selects Gribov copies with $\lambda_1$ values from
a given range. For this it may be enough to concentrate first on the less
expensive SU(2) gauge group as done here. The gluon and ghost propagators for
SU(2) and SU(3) were seen to differ only little
\cite{Cucchieri:2007zm,Sternbeck:2007ug}; hence no qualitative changes
are expected. 

One should also check how the distribution of $\lambda_1$
(\Fig{fig:eigenvalues_vs_conf}) changes on larger and finer lattices. From
\cite{Zwanziger:2003cf} and our own study in \cite{Sternbeck:2005vs}
one would expect that the distribution of $\lambda_1$ for random Gribov copies
shrinks towards smaller values the larger the lattice size.\footnote{Though note
that in \cite{Sternbeck:2005vs} the effect was not analyzed for varying lattice
spacings and a fixed lattice volume. And it would also be compatible if only
the peak (see \Fig{fig:eigenvalues_vs_conf}) shifts to lower values with
increasing volume.} If this is also the case for the
distribution of $\lambda_1$ for all Gribov copies of a single gauge
configuration, a distinction between \lc\ and \hc\ copies would be less
and less possible. On the other hand, the continuum results of
\cite{Fischer:2008uz,LlanesEstrada:2012my} and all
currently available lattice results do not indicate that in the
infinite-volume limit the Gribov problem disappears and only an infrared
diverging ghost dressing function and infrared vanishing gluon propagator
remains, as one would expect for $\lambda_1\to 0$ from
\cite{Zwanziger:2003cf}. But this certainly deserves further study.

Let us finally comment on the ``rigorous bounds'' introduced in
\cite{Cucchieri:2007rg,Cucchieri:2008fc} to control the infinite-volume
extrapolations of lattice data for the gluon and ghost propagator. Since
these bounds are composed of the lattice gluon field (for the gluon propagator),
and of $\lambda_1$ and the eigenmode~$\vec{\Phi}_1$ (for the ghost propagator),
these will suffer from similar Gribov problems as seen for the propagators here.
For a particular lattice implementation of Landau-gauge Yang-Mills theory and a
particular selection of Gribov copies these bounds will certainly constrain the
lattice data, but one should keep in mind these bounds may change when using
other copies or implementations.

Summarizing,  with the current (standard) lattice implementation of Landau-gauge
Yang-Mills theory, one seems to be able to realize varying decoupling-type
solutions for the gluon and ghost propagators within a small parameter range. 
This is possible by supplementing the lattice Landau-gauge condition by an
additional constraint, as, e.g., the one here or in \cite{Maas:2009se}. 
But it will perhaps be impossible to find an indication for the scaling
solution on a finite lattice. Even if we were able to find for every gauge
configuration that Gribov copy with the smallest $\lambda_1$, this eigenvalue
and all the others would still be non-zero, i.e., this copy is not on the
Gribov horizon. It remains to be seen if for such an approach the scaling
behavior can then appear in the limit of infinite volume and zero lattice
spacing, as discussed above.

\section*{Acknowledgments}

This work was supported by the European Union under the Grant Agreement number
IRG 256594. We thank C.~Fischer, A.~Maas and J.~Pawlowski for helpful comments
and providing partly unpublished data. We acknowledge generous support of
computing time from the HLRN (Germany).


\end{document}